\documentclass[%
 reprint,
 aps
]{revtex4-2}

\usepackage[utf8]{inputenc}
\usepackage{bbm}
\usepackage{appendix}

\usepackage[usenames,dvipsnames]{xcolor}
\usepackage{color}
\usepackage{colortbl}

\definecolor{qctrl_primary}{HTML}{680Ce9}
\definecolor{qctrl_secondary}{HTML}{BF04DC}
\definecolor{qctrl_noise}{HTML}{7B7479}
\definecolor{qctrl_axis_labels}{HTML}{514B4F}
\definecolor{qctrl_borders}{HTML}{CFCBCE}
\definecolor{qctrl_blue}{HTML}{4177D8}
\definecolor{qctrl_aqua}{HTML}{32A4A8}
\definecolor{qctrl_green}{HTML}{32A857}
\definecolor{qctrl_lime_green}{HTML}{A2A933}
\definecolor{qctrl_orange}{HTML}{D6742F}
\definecolor{qctrl_red}{HTML}{D84144}
\definecolor{qctrl_fuchsia}{HTML}{D84190}

\usepackage{blindtext}

\usepackage{amssymb,amsmath,amsthm}
\usepackage{mathtools} 

\usepackage{mathrsfs}
\usepackage{dsfont} 
\usepackage[nice]{nicefrac}

\usepackage{cancel}
\usepackage{bm}

\RequirePackage{algorithm}
\RequirePackage{algorithmic}

\theoremstyle{definition}

\theoremstyle{remark}

\usepackage{physics}
\usepackage{siunitx} 
\usepackage{qcircuit}
\usepackage{braket}

\usepackage{graphicx}
\usepackage{float}
\usepackage{rotating}

\usepackage{dcolumn}
\usepackage{multirow}
\usepackage{makecell}
\usepackage{bigdelim}
\usepackage{blkarray}
\usepackage{array}

\usepackage{hyperref}
\hypersetup{
final=true,
plainpages=false,
pdfstartview=FitV,
pdftoolbar=true,
pdfmenubar=true,
bookmarksopen=true,
bookmarksnumbered=true,
breaklinks=true,
linktocpage,
colorlinks=true,
linkcolor=blue,
filecolor=blue,      
urlcolor=cyan,
citecolor=blue,
anchorcolor=green
}

\usepackage[nameinlink]{cleveref}
\crefname{equation}{Eq.}{Eqs.}
\crefname{align}{Eq.}{Eqs.}
\crefname{figure}{Fig.}{Figs.}
\crefname{table}{Table}{Tables}
\crefname{tabular}{Table}{Tables}
\crefname{section}{Sec.}{Secs.}
\crefname{appendix}{App.}{Apps.}

\crefname{appsec}{App.}{Apps.}
\crefname{appchapter}{App.}{Apps.}

\crefname{algorithm}{Algo.}{Algos.}
\creflabelformat{equation}{#2#1#3}

\usepackage{natbib}
\usepackage{diagbox}
\begin{document}

\title{Experimental Deep Reinforcement Learning for Error-Robust Gateset Design on a Superconducting Quantum Computer}

\author{Yuval Baum}
\author{Mirko Amico}
\author{Sean Howell}
\author{Michael Hush}
\author{Maggie Liuzzi}
\author{Pranav Mundada}
\author{Thomas Merkh}
\author{Andre R. R. Carvalho}
\author{Michael J. Biercuk}
\altaffiliation[Also ]{ARC Centre for Engineered Quantum Systems, The University of Sydney, NSW Australia}
\affiliation{%
 Q-CTRL, 
 Sydney, NSW Australia \& Los Angeles, CA USA
}%

\date{\today}


\begin{abstract}

Quantum computers promise tremendous impact across applications -- and have shown great strides in hardware engineering -- but remain notoriously error prone.  Careful design of low-level controls has been shown to compensate for the processes which induce hardware errors, leveraging techniques from optimal and robust control.  However, these techniques rely heavily on the availability of highly accurate and detailed physical models which generally only achieve sufficient representative fidelity for the most simple operations and generic noise modes. In this work, we use deep reinforcement learning to design a universal set of error-robust quantum logic gates on a superconducting quantum computer, without requiring knowledge of a specific Hamiltonian model of the system, its controls, or its underlying error processes.  We experimentally demonstrate that a fully autonomous deep reinforcement learning agent can design single qubit gates up to $3\times$ faster than default DRAG operations without additional leakage error, and exhibiting robustness against calibration drifts over weeks.  We then show that $ZX(-\pi/2)$ operations implemented using the cross-resonance interaction can outperform hardware default gates by over $2\times$ and equivalently exhibit superior calibration-free performance up to 25 days post optimization using various metrics.  We benchmark the performance of deep reinforcement learning derived gates against other black box optimization techniques, showing that deep reinforcement learning can achieve comparable or marginally superior performance, even with limited hardware access. 
\end{abstract}

\maketitle



\section{Introduction}
Large-scale fault-tolerant quantum computers are likely to enable new solutions for problems known to be hard for classical computers. The quantum information community has recently made great strides towards realizing such systems;
a significant step towards quantum advantage (when a quantum computer can solve a practically relevant problem ``faster'' than a classical computer) was demonstrated by Google~\cite{AruteNature2019} and the Chinese Academy of Sciences~\cite{Zhong1460}; and cloud-based quantum computing offerings are now commercially available~\cite{IBM_website, Amazon_website, Microsoft_website, Rigetti_website}. However, demonstrating a computational advantage on a problem of practical importance remains an outstanding challenge for the sector. 

The extreme sensitivity of today's hardware to noise, fabrication variability, and imperfect quantum logic gates are the key factors limiting the community's ability to reliably perform quantum computations at scale~\cite{Preskill2018}. Fortunately, it has been repeatedly demonstrated that the use of robust and optimal control techniques for gateset design may lead to dramatic improvements in hardware performance and downstream computational capabilities, as a complement to both ongoing hardware improvements and future application of quantum error correction \cite{Huang1983, CLARK2003, NURDIN2009, biercuk2009, Yale_Optimal, carvalho2020errorrobust, DongIET2010, Kofman2004, Gordon2008, Yao2007, Khodjasteh2005, Byrd2003, Viola2003, Vitali1999, Viola1998, Chaudhury2007, Jessen2007, Motzoi2009, Yale_Optimal,  soare2014, werninghaus2020leakage, leng2019robust, Milne_MS_2020, ball2020software, Wittler2021}. The design process is straightforward in cases where Hamiltonian representations of both the physical and the noise models in the underlying system are precisely known, but has proven considerably more difficult in state-of-the-art large-scale experimental systems. A combination of effects introduces challenges not faced in simpler systems including: unknown and transient Hamiltonian terms~\cite{Magesan2020} with nonlinear dependence on applied control signals~\cite{Reagor_2018}; control signal distortion in transmission~\cite{Rol_2020}; undesired crosstalk~\cite{Sheldon2016} and frequency collisions between neighboring qubits; and time varying environmental noise. In all cases, complete characterization of Hamiltonian terms~\cite{Sheldon2016}, their functional dependencies, and their temporal dynamics becomes unwieldy as the system size grows.

In this manuscript, we overcome this fundamental challenge by demonstrating the experimental use of Deep Reinforcement Learning (DRL) for the efficient design of an error-robust universal quantum-logic gateset on a superconducting quantum computer without any \textit{a priori} hardware model. 
In our approach, we design an agent which iterative and autonomously constructs a model of the \textit{relevant} effects of a set of available controls on quantum computer hardware, incorporating both targeted responses and undesired effects such as cross-couplings, nonlinearities, and signal distortions. We task the agent with learning how to execute high fidelity constituent operations which can be used to construct a universal gateset -- an $R_{X}(\pi/2)$ single-qubit driven rotation and a $ZX(-\pi/2)$ multiqubit entangling operation -- by allowing it to explore a space of piecewise constant (PWC) operations executed on a superconducting quantum computer programmed using Qiskit Pulse~\cite{Qiskitpulse_2020}, and with constrained access to measurement data in stark contrast to previous theoretical studies~\cite{Boixo_RL}. First, we demonstrate that single-qubit gates can be designed which outperform the default DRAG gates in randomized benchmarking with up to a $3\times$ reduction in gate duration, down to $\approx 12.4$~ns, and without evident increases in leakage error.
Due to the training process, these gates are shown to exhibit robustness against common system drifts (characterized in~\cite{carvalho2020errorrobust}), providing up to two weeks of outperformance with no need for intermediate re-calibration of the gate parameters; the daily performance variation of these gates is consistent with limits imposed by measured fluctuations in device $T_{1}$. Next, we show that the DRL agent is able to create novel implementations of the cross-resonance interaction which show up to $\approx 2.38\times$ higher fidelity than calibrated hardware defaults. We characterize gate performance using both interleaved randomized benchmarking and gate repetition to better reveal coherent errors, achieving $\mathcal{F}_{ZX}>99.5\%$ across multiple hardware systems, and maintaining $\mathcal{F}_{ZX}>99.3\%$ over a period of up to 25 days with no additional re-calibration. Finally, we demonstrate the use of these DRL defined entangling gates within quantum circuits for the SWAP operation and show $1.45\times$ lower error than calibrated hardware defaults. Across these demonstrations we benchmark the DRL designed two-qubit gates against gates created using a black box automated optimization routine leveraging a custom implementation of Simulated Annealing (SA) and observe comparable performance between the two methods, suggesting incoherent processes as the current bottleneck.  

\section{Optimized Quantum Logic Design with Deep Reinforcement Learning}\label{gate_op}
We begin by defining the problem of quantum logic gate optimization and its measures of success. Consider a conventional Hamiltonian description for coupled transmons written using a control Hamiltonian, $H_{\text{ctrl}}(t)\left[\{\Omega^\omega_{I, j}(t), \Omega^\omega_{Q, j}(t)\}\right]$.  This Hamiltonian possesses some functional dependence on applied time varying microwave control signals $\Omega_{I/Q, j}^\omega$ targeting qubits labeled by index $j$ and written in a conventional $I/Q$ decomposition for a drive at frequency $\omega$.  All other terms that are not in our control Hamiltonian are contained in an additional term $H_{\text{system}}(t)$, including the bare qubit Hamiltonian, deterministic ``drift'' terms and stochastic noise terms.  Together we have $H_{\text{tot}}(t)=H_{\text{ctrl}}(t)+H_{\text{system}}(t)$.

The aim of the control problem is to find the optimal set of functions $\{\Omega^\omega_{I, j}(t), \Omega^\omega_{Q, j}(t)\}$, such that the resultant unitary evolution, $\mathcal{U} = \mathcal{T} \exp\big(-i\int_0^T\,dt'H_{\text{tot}}(t')\big)$, matches a target unitary $\mathcal{U}_{\text{target}}$. In the above, $T$ is the total evolution time and $\mathcal{T}$ is the time-ordering operator which defines the calculation procedure. 
Typically, numerical optimization~\cite{Byrd1995, KHANEJA2005, Zhu1997, carvalho2020errorrobust} techniques may be employed in appropriate reference frames in order to construct the controls via minimization of the gate infidelity
\begin{equation} \label{Eq:err}
\mathcal{I} = 1 - \mathcal{F} = 1 - \frac{1}{D^2}\left|\textrm{Tr}\left(\mathcal{U}^{\dagger}\mathcal{U}_{\text{target}}\right)\right|^2.
\end{equation}
for a $D$-dimensional Hilbert space.
Performing useful gate optimization in this way -- especially when trying to exceed state-of-the-art experimental fidelities in realistic systems -- requires a comprehensive understanding of the relevant terms in the system Hamiltonian.

In practice, the real Hamiltonian of the system typically may include nonlinear distortions of the control fields ($F_{I/Q}(\cdot)$), nonlinear coupling of the distorted control fields into new Hamiltonian terms ($H^{\text{nl}}_{\text{ctrl}}$), and additional hidden terms in the Hamiltonian that may change with application of the control fields:
\begin{align*}\label{the_model}
\Tilde{H}_{\text{tot}}(t) \to 
& H_{\text{ctrl}}(t)\big[F_{I/Q}(\Omega_{I/Q, j}^\omega(t)),t \big]+ \\
& H^{\text{nl}}_{\text{ctrl}}\big[ F_{I/Q}(\Omega_{I/Q, j}^\omega(t)),t \big] + \\
& H_{\text{hidden}}(t) + H_{\text{system}}(t)
\end{align*}

It is not generally tractable to identify this diverse range of Hamiltonian terms in large interacting systems, and even small simplifying approximations will typically lead to catastrophic failure of numerically optimized controls due to the sensitive manner in which the system is steered through its Hilbert space.

An alternative approach to quantum logic design based on iterative interactive learning obviates the need of having an accurate representative model of the physical system -- in particular the various new terms appearing in $\tilde{H}_{\text{tot}}$. Deep reinforcement learning techniques stand out in their ability to deal with high-dimensional optimization problems and in the absence of labeled data or an underlying noise model~\cite{sutton2018reinforcement, Recht2019}. In DRL, an agent interacts with its environment by taking actions and receiving feedback in the form of state observables and rewards. By learning to maximize these rewards, the agent learns a targeted behavior such as coordinated robotic motion~\cite{kormushev2013reinforcement}, autonomous driving~\cite{sallab2017deep}, or in the present case, how to perform high fidelity quantum logic operations. The uniqueness of DRL comes from the fact that, unlike in other closed-loop optimization methods, intermediate information is extracted as well as a final measure of reward in order to construct and refine an internal model of the system's most relevant dynamics (this model need not be interpretable under human examination). 

\begin{figure} 
\centering
\includegraphics[width=\linewidth]{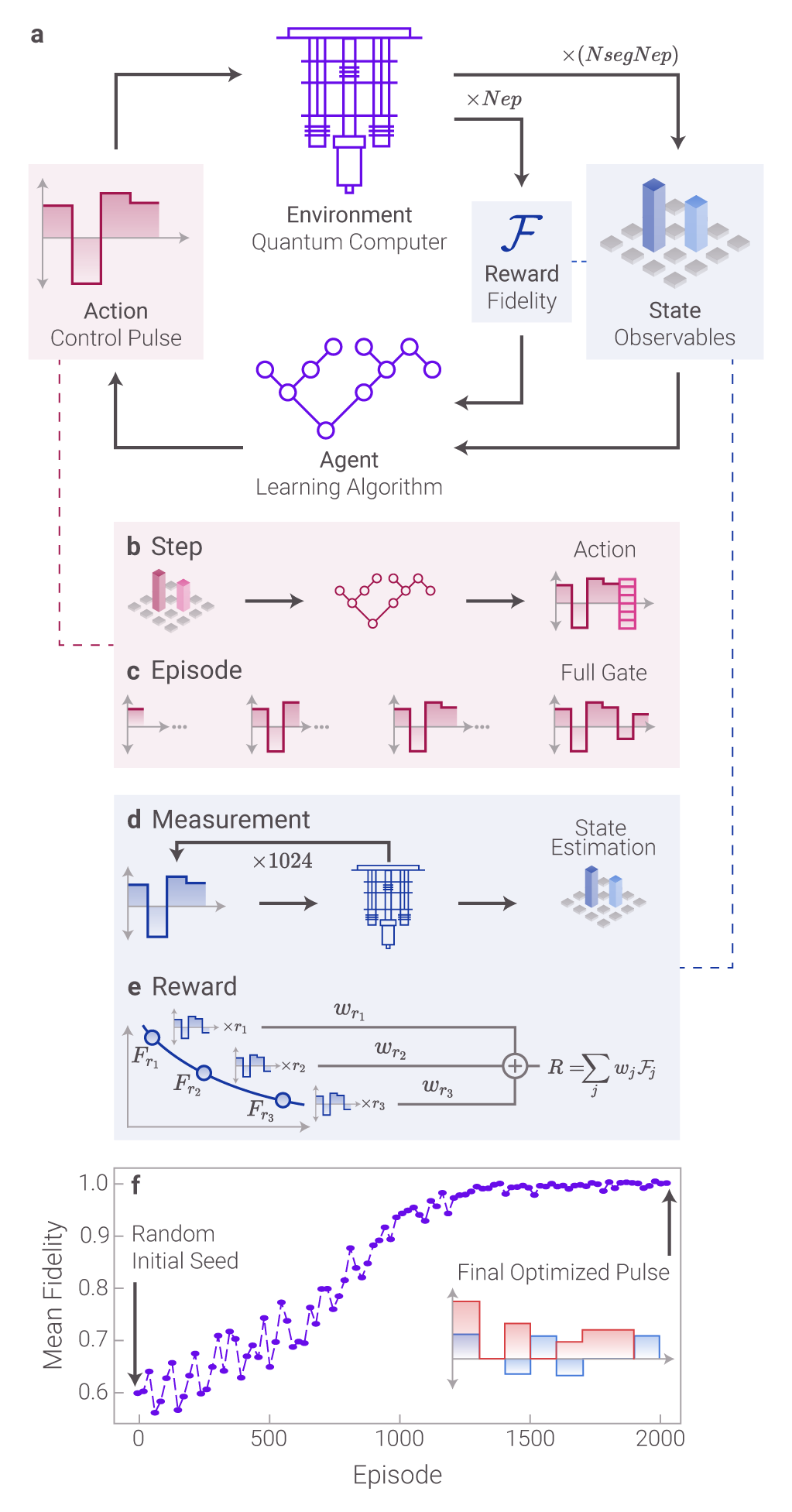}
\caption{DRL optimization on quantum hardware. 
(a) In the DRL cycle, an intermediate information is collected in order to optimize a long term reward function. The left side of the loop indicates actions taken by the agent and the right captures measurements returned
(b) At each DRL step the action, \textit{e.g.}, amplitude and phase of the next segment(s) is chosen from a set of allowed actions while the previous segments actions are held fixed. 
(c) A set of actions that form a full control pulse is an episode. 
(d) After every step, an estimation of the quantum state is performed using simplified tomography, and at the conclusion of each episode an estimate of fidelity is made. Each constituent measurement contributing to this process is repeated 1024 (256) times for final (intermediate) states. In order to reduce the effect of the readout errors, we employ a standard measurement error mitigation scheme~\cite{IBMErrorMitigation}. 
(e) In order to estimate the gate quality, a reward is constructed by performing a weighted average of measured fidelities for sequences employing different numbers of control pulse repetitions.(f) Example of DRL optimization convergence for a control pulse implementing a single-qubit gate.}
\label{fig:rl_scheme}
\end{figure}

The basic ingredients of DRL are \textit{states}, \textit{actions} and \textit{rewards} (Fig.~\ref{fig:rl_scheme}a). 
The \textit{state} is snapshot of the \textit{environment} at a given time, and in the context of gate optimization, can be any suitable representation of the quantum state of the quantum device. The \textit{action} is the means by which we affect or stir the state of the environment. For example, an action can be a low level electromagnetic control pulse or any other operation that alters the state of the environment. Finally, the \textit{reward} encapsulates the feedback from the environment.  After an action is executed in the environment, the state of the environment is observed again, and a reward is given in order to quantify the quality of the last action. The essence of DRL is the ability to estimate and maximize the long term cumulative reward learned when a deep neural network based agent interacts with an environment through many trials.

DRL has recently been deployed for a variety of quantum control problems using numerical simulation~\cite{ZhengQENG2020}, including both theoretical gate optimization~\cite{Boixo_RL,an2019, NiuNPJ2019} and other tasks~\cite{BukovPRX2018, PorottiPhysComm2019, ZhangNPJ2019, ZhengPhysRevA2021, pozzi2020using, LamataSR2017, KhalilGov2020, borah2021measurement, sivak2021modelfree}. In these simulation based studies it was possible to make at least one of the following strong assumptions: the system suffers zero noise or has a deterministic error model; controls are perfect and instantaneous; and quantum states are completely observable across time.
Moving beyond these studies to efficient experimental implementations on real quantum computers, we focus on designing DRL algorithms compatible with realistic execution and measurement constraints and the complexity of $\tilde{H}_{\text{tot}}$.

This involves overcoming three main challenges that we discuss below: (i) creating an efficient representation of the effective Hamiltonian in a manner compatible with experimental implementation, (ii) creating a suitable measurement routine compatible with the limited observability of the system state and its unitary evolution in a quantum computer, (iii) designing appropriate agents which can efficiently learn system dynamics based on these constrained controls and limited access to measurements.  An overview of the complete DRL learning process we employ to perform experimental gateset design using DR is featured in Fig.~\ref{fig:rl_scheme}, and complemented by pseudocode in Alg.~\ref{algo:1}. 

First, we seek an effective PWC control Hamiltonian, which physically corresponds to control pulses with a PWC envelope. The $N_{seg}$-segment control Hamiltonian may be written as
\begin{equation}\label{eq:ctrl_Ham}
    H_\text{ctrl}(t) = \sum\limits_{k=1}^{N_{seg}} H_{k} \mathds{1}_{k}(t),
\end{equation}
where $\mathds{1}_{k}(t) = 1$ if $(k-1)\Delta t<t< k\Delta t$ and is zero otherwise, with $\Delta t$ being the duration of each segment.  The definition of this PWC control Hamiltonian -- which can have multiple constituent terms -- is found through the iterative reinforcement learning process shown schematically in Fig.~\ref{fig:rl_scheme}b-c.  This choice is convenient due to limitations in programming and manipulating real hardware, and any deviations from the idealized PWC waveform applied to the device, due to \textit{e.g.} signal distortions, are directly captured in the measurements performed by the agent and the effective model it constructs.  

Next, in order to effectively learn the hardware model, we observe and store the state of the quantum computer in intermediate steps. We observe the state after step $k$, and allow the agent to choose the action for the next step. Due to quantum state collapse on projective measurement, at the beginning of a new step our protocol must first reinitialize the qubits and repeat the exact sequence of actions through step $k$, before applying the new action at step $k+1$.  Thus we are able to sequentially build up to a complete execution of a candidate quantum logic gate at the conclusion of an episode over $N_{seg}$ state-observation cycles (giving a total of $N_{seg}N_{ep}$ state observations over $N_{ep}$ episodes).  

The measurement protocols we employ to observe the system state are based on the concepts of quantum state tomography (Fig.~\ref{fig:rl_scheme}d). For optimization of the single-qubit $R_{X}(\pi/2)$, we prepare and measure qubits in the three different Pauli bases, and also measure population leakage beyond the computational subspace of the transmon qubit. For the two-qubit $ZX(-\pi/2)$ entangling gate, we perform full tomography of the computational basis by collecting nine measurements in order to evaluate the expectation values of all Pauli strings. The measurement protocol is repeated to build projective-measurement statistics, and several different initial states are chosen in order to specify the gate uniquely. For all gates the state of the system is represented by a real vector of expectation values. We find that this approach provides a sufficiently reliable approximation of the state and system dynamics.  
In addition to state observation we must explicitly calculate the reward at the end of an episode.  To do this the fidelity is estimated as an element of the reward at the end of full episodes, \textit{i.e.}, after the full implementation of the candidate gate, and thus occurs with the number of episodes, $N_{ep}$. We evaluate fidelity relative to the target operation by acting on each initial quantum state of the qubits with a variable number of gate repetitions. The final fidelity of the waveform is then estimated as a weighted mean of the different repetitions applied to the different initial states (Fig.~\ref{fig:rl_scheme}e). The set of initial states and the number of gate repetitions are chosen such that the gate operation is uniquely tested and the cost/reward function captures the theoretical error in Eqn.~\ref{Eq:err}. This repetition-based measurement scheme serves to amplify the gate error in order to overcome the so-called state preparation and measurement (SPAM) error endemic to real hardware, and estimated at $\approx 4 \%$ in the hardware we employ. Combining fidelity estimates produced for different numbers of gate repetitions also averages over pathological contextual errors that arise in experimental hardware as circuit lengths vary.

\begin{figure}[t!]
\begin{algorithm}[H] 
\caption{\hspace{6px} DRL Training Loop} 
\label{algo:1}
\begin{algorithmic}[1]
 \REQUIRE Initial policy parameters $\theta$\\
  \FOR {episode $= 1,2,\ldots$}
   \FOR {each initial state $j= 1,2,\ldots$}
   \STATE Choose first action according to $\pi_\theta(a_{0} | \mathbf{s}_{0})$ \COMMENT{$\mathbf{s}_{k}=$ state tomography after $k$ steps}\\
    \FOR {$k = 1,2,\dots,N_\text{seg}$}
     \STATE Initialize the quantum state of the qubit(s) $|\psi_{0, j}\rangle$\\
     \STATE Evolve the state by the first $k$ segments\\
     \STATE Estimate the qubit(s) state $\mathbf{s}_{k}$\\
     \STATE Choose next action according to $\pi_\theta(a_{k} | \mathbf{s}_{k})$\\
     \STATE Store trajectory $(\mathbf{s}_{{k-1}},a_{{k-1}},\mathbf{s}_{k})$\\
 	\ENDFOR
 	\FOR {$p$ in repetitions}
 	 \STATE Repeat the final pulse $p$ times\\
 	 \STATE Measure, compute and store state fidelity\\
    \ENDFOR
   \ENDFOR
   \STATE Compute reward based on state fidelities\\
   \STATE Update the policy's parameters $\theta$\\
  \ENDFOR
 \end{algorithmic} 
\end{algorithm}
\end{figure}

Finally, we design a DRL algorithm which maximizes the long term reward using an agent compatible with the above constraints, based on a policy gradient optimization algorithm. A policy $\pi(a|s)$ is a function that receives as an input the state of the system and returns a probability distribution over actions. This distribution is then used to decide the next action such that over many steps and episodes the reward is maximized. The policy function $\pi(a|s)$ is represented by a deep neural network whose trainable parameters are updated during the learning process in order to efficiently approximate the optimal policy over this space.  A rigorous comparison between the performance of different DRL algorithms for this problem in a simulated environment -- which led to our selection of the policy gradient -- was investigated in~\cite{Liuzzi2021}. 

\section{Experimental Deep Reinforcement Learning on Superconducting Quantum Computers}
\begin{figure*}[t] 
\centering
\includegraphics[width=0.97\linewidth]{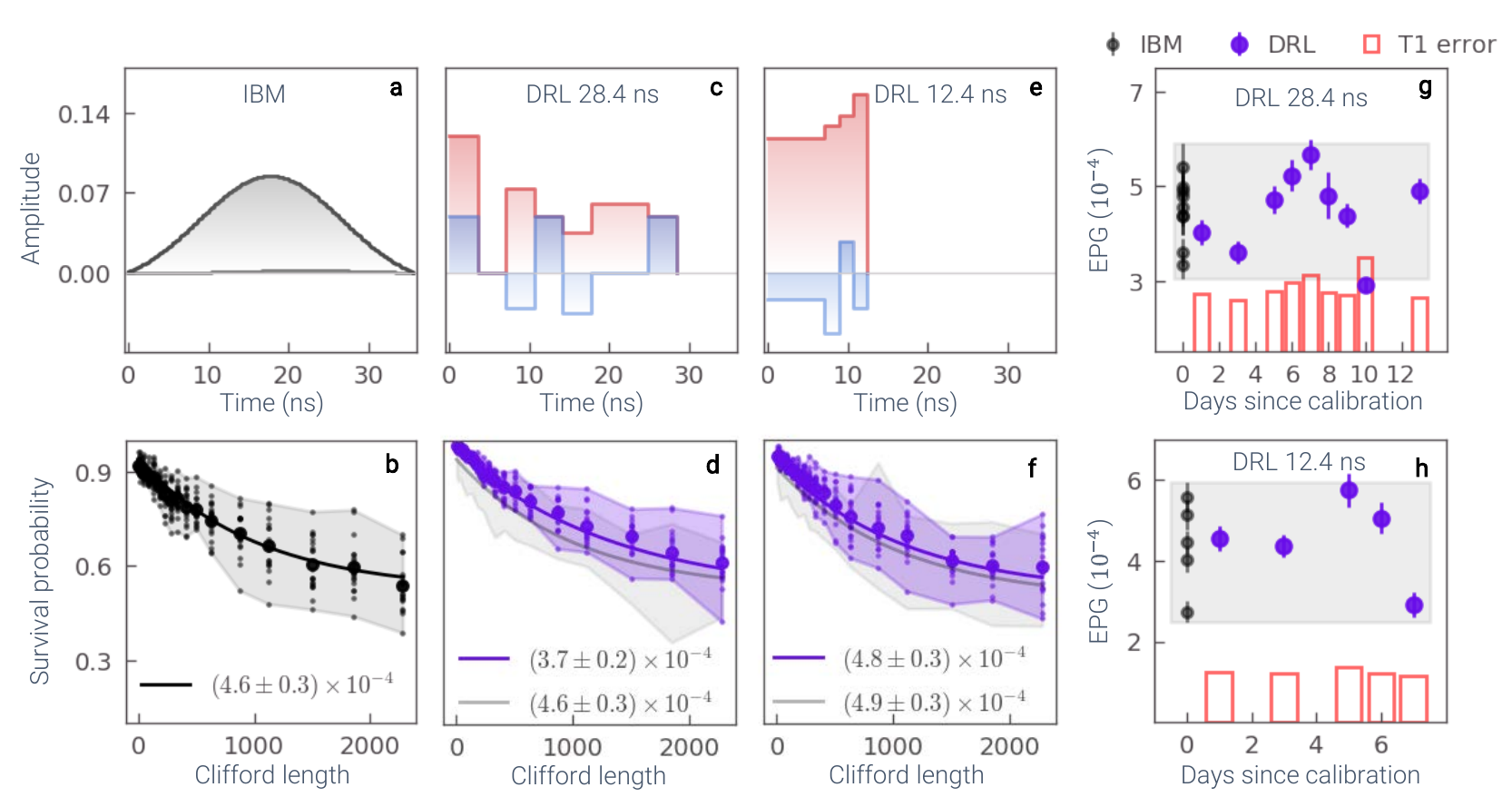}
\caption{$R_x(90)$ DRL optimization on the IBM device Rome. (a) The pulse waveform for the IBM default $36$ ns pulse, and (b) associated gate error estimation via randomized benchmarking. Here, shading indicates the spread of individual sequence survival probabilities and an exponential fit is produced to the mean of the distribution for each sequence length. (c-d) Results of DRL optimization for a $28.4$ ns pulse, and  (e-f) an ultra short $12.4$ ns pulse. Both report minor improvements in estimated gate fidelity (see text for discussion).  In panels (d) and (f) only the best fit decay curve and shading over individual sequences is shown for clarity.
(g-h) RB-based demonstration of robustness for DRL optimized pulses.  Black markers show the RB performance of the daily calibrated default and colored markers indicate (g) the $28.4$ optimized pulse and (h) the $12.4$ optimized pulse.  Both DRL optimized pulses were defined on day zero and repeated without additional calibration on subsequent days.  For each day an estimation of the $T_1$ contribution to the error is plotted using red bars, based on the tabulated $T_{1}$ provided by IBM. On a daily basis the DRL pulses are up to $25 \%$ better than the daily calibrated IBM pulses and performance fluctuations closely track the daily performance of the IBM pulses and $T_{1}$ limits.  DRL pulses remain within a band of natural hardware fluctuations near the $T_{1}$ limit up to two weeks past gate definition.}
\label{fig:1q_rl_results}
\end{figure*}

The DRL algorithm is executed on experimental hardware via cloud-access to a superconducting quantum computer operated by IBM. Commands are executed using Qiskit Pulse to program various accessible analog control channels relevant to implementation of single and multiqubit gates. The DRL agent is separately hosted on a cloud server in order to allow an efficient learning procedure and is provided command of the quantum computer for fixed blocks of time. 

The primary experimental constraint we face is limited hardware access, and our approach must function even with these restrictions. In order to reduce the effect of overhead due to cloud access to hardware, we generally batch several episodes in the learning procedure, meaning we execute them sequentially prior to the resulting measurement data being provided to the agent.  With our selected DRL-agent implementation, convergence typically occurs over as little as $10-20$ experimental batches (batch sizes for single and two qubit gates are $25$ and $16$ respectively), which corresponds to $0.5-1$ wall-clock hours. This time is dominated by hardware-API and access-queue times, with typical total experimental execution of less than a minute, and agent calculations consuming negligible time on a cloud server. 

An example optimization convergence over episodes for a single-qubit gate (see details below) is shown in Fig.~\ref{fig:rl_scheme}f, and notably requires more than an order of magnitude less episodes than previous numerical studies~\cite{Boixo_RL} to achieve a high fidelity gate. The convergence need not exhibit a monotonic increase in fidelity as the DRL agent is allowed to freely explore the space of available controls. Once the learning process converges, the resultant gates are consistently nontrivial, showing structure in the relevant control parameters but exploiting physics which is not obvious upon examination.  

We now describe the specific optimizations we have performed and benchmark the performance of the resulting gates.  Beginning with the single-qubit gate we target optimization of $R_x(\pi/2)$, a $\pi/2$ radian rotation around the $x$ axis. This gate is implemented as a driven, microwave mediated operation and serves as a fundamental building block for arbitrary single-qubit rotations in a $U3$ decomposition when combined with virtual $Z$-rotations. In the following, all pulses are presented in terms of arbitrary amplitudes for the $I$ and the $Q$ components of each driving channel.

We target gates with reduced duration and enhanced error-robustness relative to a default $36$ ns analytic DRAG~\cite{Motzoi2009} pulse calibrated through a daily routine that is inaccessible to us.  We select a gate duration and task the DRL agent with discovering high fidelity PWC waveforms with eight time segments, constituting a total 16-dimensional optimization when accounting for freedom in both the $I$ and $Q$ channels. Again due to the iterative and stepwise nature of DRL algorithms, the effective initial seed is random.  

\begin{figure*}[t!] 
\centering
\includegraphics[width=0.98\linewidth]{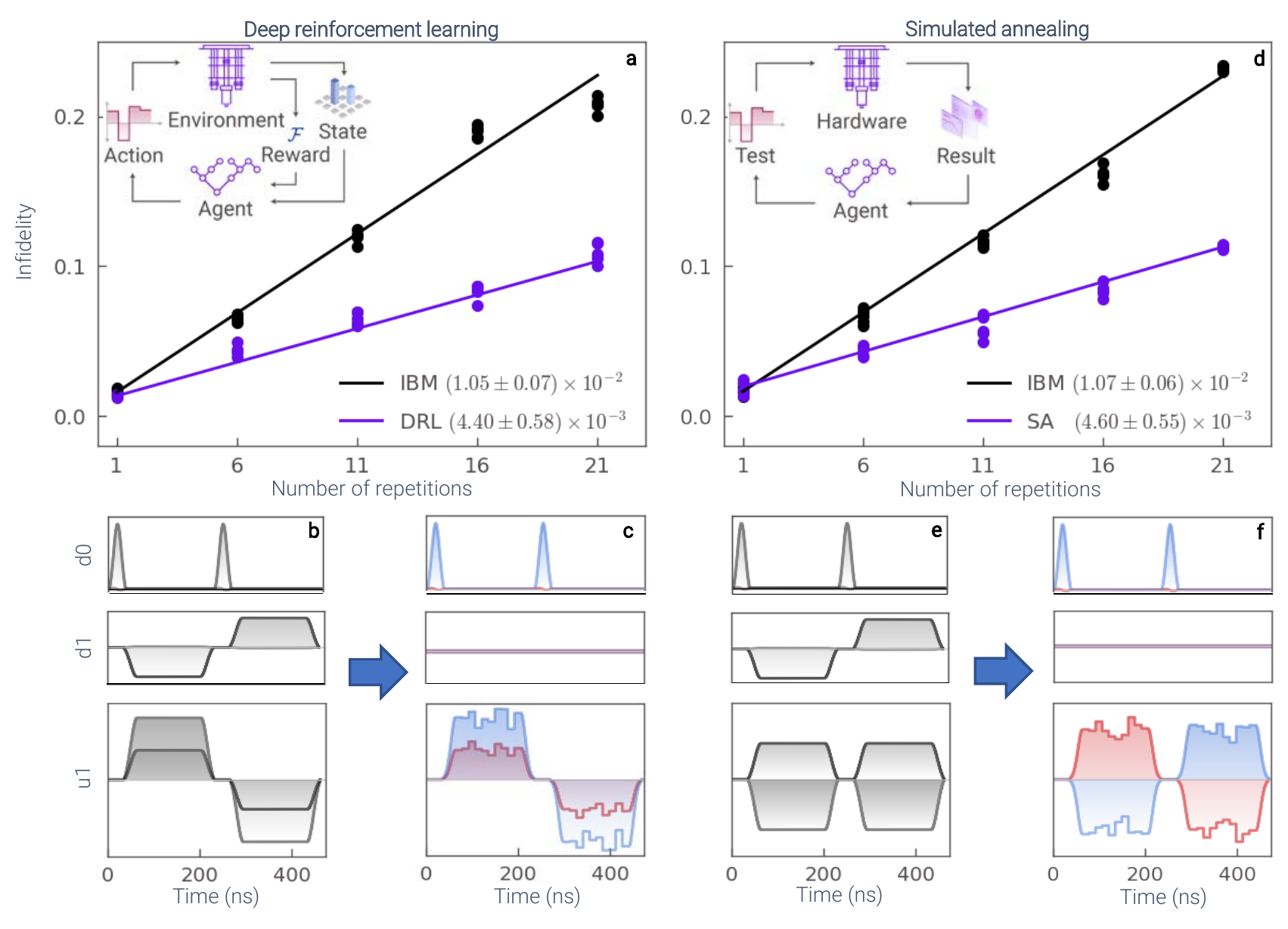}
\caption{Optimization of multiqubit $ZX(-\pi/2)$ entangling gates on IBM device Casablanca, benchmarked against a black box closed-loop optimization routine.  
(a) Infidelity measurement for circuits consisting of different numbers of gate repetition for the IBM default and DRL optimized gate. For each number of repetitions we perform $5$ experiments on each of two different initial states $|00\rangle$ and $|10\rangle$. Approximate gate error is extracted from the slope of infidelity with gate repetition.  (Upper Inset) Schematic of the optimization cycle.
(b-c) The programmed control waveforms for the (b) IBM default and (c) DRL optimized $ZX(-\pi/2)$ gates. Note that channel d1, used as a cancellation tone in the IBM default is not used in the optimized gate. 
(d-f) Similar plots to (a-c) as achieved via simulated annealing (SA) on the same IBM hardware. Derived benefit using SA is similar to DRL, showing $200\%$ improvement over the IBM default gate. These initial performance calibration measurements were performed $\approx 48$ h after initial gate design due to hardware access constraints and comparison is made to the most recently updated default gate.}
\label{fig:2q_gates}
\end{figure*}

We select two target gate durations informed by hardware constraints to serve as illustrative examples.  First, we choose a PWC pulse with a total duration $28.4$ ns, $20\%$ shorter than the default, because the default gate performance is near the $T_{1}$ limit and leaves only approximately $20-30\%$ maximum achievable performance enhancement in base gate fidelity. Next, we select a gate which is $12.4$ ns, or $\approx 3\times$ shorter than the default in order to probe the ability of the RL agent to suppress leakage arising from fast pulses with spectral weight overlapping higher-order transitions. Further details on the reward/cost function in use are presented in the Appendix.   
The results of DRL gate optimization executed on a superconducting quantum computer called \textit {ibmq\_rome} are shown in Fig.~\ref{fig:1q_rl_results}. We evaluate the performance of the gate implementation by utilizing Clifford randomized benchmarking (RB) \cite{Knill2008} which provides an estimate of average error-per-gate (EPG).  The $24$ Clifford gates used in RB are generated using the $R_x(\pi/2)$ gate together with virtual $Z$-rotations in a $U3$ decomposition, and sequences are constructed using a custom compiler allowing incorporation of arbitrary gate definitions into RB sequences (see Appendix A for details). 

In Fig.~\ref{fig:1q_rl_results} we see that the $28.4$ ns optimized pulse achieves an EPG $3.7\times 10^{-4}$, $\approx25 \%$ lower than the default and consistent with expectations based on $T_{1}$ limits.  Further, we observe reduced variance of individual sequence performance about the mean (indicated by colored shading), consistent with additional suppression of coherent errors~\cite{Ball_2016, Mavadia_2018, carvalho2020errorrobust}. We have observed performance enhancements $\approx 2.13\times$ in RB using $28.4$ ns gates defined via a black box closed-loop optimization; at this stage we are not able to distinguish whether this difference is due to the underlying method of gate optimization or the fact that a different machine was employed for these tests (see Appendix C).

\begin{figure*}[t] 
\centering
\includegraphics[width=\linewidth]{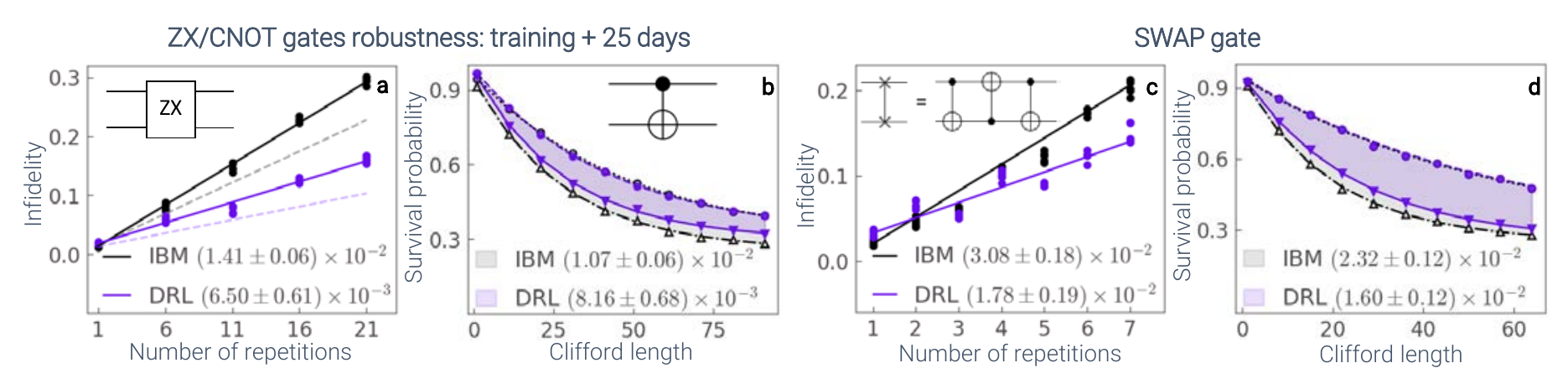}
\caption{Robustness and circuit level implementations of DRL optimized two-qubit gates. (a) Repetition measurements taken immediately after optimization (dotted lines) and 25 days following optimization. The default gate has been recalibrated but the DRL optimized gate remains unchanged over 25 days. (b) Interleaved Randomized Benchmarking (IRB) comparisons at 25 days post-DRL optimization. Here shading represents the gap between the overall sequence and the CNOT constructed from the default or optimized $ZX(-\pi/2$). The reduced purple shaded area indicates improved performance relative to the default. (c) Repetition and (d) IRB for a DRL optimized SWAP gate constructed from 3 CNOTS (inset, panel c).}
\label{fig:swap}
\end{figure*}

The performance of the $12.4$ ns optimized pulse is comparable to the default, despite being $3\times$ faster, indicating that leakage errors can be suppressed via appropriate definition of the DRL reward function.  It is not clear whether the lack of further improvement arises due to an overly strict constraint on the actions afforded to the agent, or whether there is a trade off between increased leakage errors and reduced incoherent errors.  

We examine the \textit{robustness} of both DRL defined gates by comparing the achieved EPG from RB for the same gates applied on different days. 
The default gates are recalibrated daily and can show amplitude variations on the order of several percent; fixed waveforms are used in each experiment involving DRL defined gates without recalibration. 
For both gates we observe that we achieve consistent performance relative to the default gates over a period up to two weeks, with measured EPG closely tracking fluctuations in the measured hardware $T_{1}$ (Figs.~\ref{fig:1q_rl_results}g, h).  In previous experiments we observed that default gates on comparable hardware could vary substantially in performance after $\approx 12$ hours elapsed since last calibration~\cite{carvalho2020errorrobust}.  These findings suggest that while temporal robustness was not explicitly included in the reward function employed, the agent may have discovered robustness as the underlying hardware varied during the training process.

For the two-qubit gate, we implement the $ZX(-\pi/2)$ gate using an entangling cross-resonance pulse \cite{Paraoanu2006, DeGroot2010, Rigetti2010, Chow2011, Chow2012, Tripathi2019, Magesan2020, Malekakhlagh2020} on the control qubit in combination with multiple single-qubit gates applied to both the control and the target qubits in an echo-like sequence ~\cite{Sheldon2016, Sundaresan2020, Patterson2019}.   The default gate implementation applies a ``square-Gaussian'' cross-resonance pulse and a simultaneous cancellation tone applied to the resonant drive of the target qubit in order to compensate for direct classical crosstalk (Fig.~\ref{fig:2q_gates}b).  

We employ the same base structure and ask the DRL agent to find $10$-segment PWC waveforms for the cross-resonance interaction \textit{without} application of an additional crosstalk cancellation tone.  This corresponds to a 20-dimensional optimization problem due to the variable amplitude and phase of the cross-resonance drive.  In this instance we also compare the DRL procedure, which builds a gate from scratch, to a black box gate optimization using an autonomous simulated annealing (SA) algorithm, and seeded with the initial calibrated default gate.  

Optimized $ZX(-\pi/2)$ gates were found using both DRL and SA on two different IBM devices. Optimizations and comparisons to the default were performed on different days (due to access limits), resulting in the variation between calibrated default gate definition and performance observed. Due to the collection of intermediate information, a typical DRL optimization step is approximately $2\times$ longer than a SA optimization step, but we observe that both optimization methods converge in roughly the same number of iterations.

We first compare gate performance using a repetition scheme in which the same gate is applied multiple times and on different initial states.  For a given number of repetitions we act with the gate on two orthogonal initial states five times each and average the state fidelity of these $10$ different experiments. From the fidelity decay with repetition number we can simply extract a gate error from the approximate slope of these curves.  

Results are summarised in Fig.~\ref{fig:2q_gates} showing that with both methods, the optimized pulses outperform the default pulses with up to $2.38\times$ reductions in error-per-gate, achieving a gate fidelity $>99.5\%$. These results show that both agents are able to identify superior $ZX(-\pi/2)$ gates without the need for use of a cancellation tone on channel $d1$ (Fig.~\ref{fig:2q_gates}c, f), and that we are able to avoid the potential pitfalls of the learning procedure using DRL relative to a direct fidelity optimization

The benefits of using DRL for the design of entangling gates can expand beyond direct improvements in instantaneous gate fidelity.  In a manner similar to the single-qubit robustness studies, we have seen that DRL optimized $ZX(-\pi/2)$ gates outperform the default even up to 25 days since optimization by $\gtrsim 2\times$, again with no recalibration or tuning (Fig.~\ref{fig:swap}a). As another measure, we construct a CNOT gate from the optimized $ZX(-\pi/2)$ gate and compare it to the default CNOT using interleaved randomized benchmarking (IRB)~\cite{Magesan2012}. The absolute IRB gate fidelities achieved and the relative improvements $\approx 25-70\%$ vary with machine in use and time (as $T_{1}$ can fluctuate substantially), but optimized gates consistently outperform the default across multiple metrics and over long delays since calibration. For the example of testing 25+ days post calibration shown in Fig.~\ref{fig:swap}b, we achieve a DRL optimized CNOT gate fidelity $>99\%$.  

Finally, we demonstrate that DRL can be used to directly optimize the SWAP gate in situ. The SWAP gate involves sequential application of three CNOT gates, built in turn from $ZX(-\pi/2)$ entangling operations and single-qubit unitaries. We maintain this overall decomposition but create a new reward function for the DRL algorithm as follows. We apply the full SWAP schedule with varying repetition values $r_{i}$ on initial states $|+0\rangle$ and $|+1\rangle$, and average the different state fidelities which we extract from a full state tomography. Again we are able to see improvements in the SWAP gate fidelity through both direct repetition and interleaved randomized benchmarking. Using DRL optimization on \textit {ibmq\_bogota}, we achieve up to $1.45\times$ improvement in the achieved interleaved randomized benchmarking fidelity.

\vspace{-12px} 
\section{Conclusions and Outlook}\label{summary}
In this work, we have shown the benefits of using deep reinforcement learning for the autonomous experimental design of high fidelity and error-robust gatesets on superconducting quantum computers. We demonstrated that by manipulating a small set of accessible experimental controls, such as the envelope functions for microwave pulses, our method was able to provide low level implementations of novel quantum gates based only on measured system responses without requiring any prior knowledge of the particular device model or its underlying error processes. 
These gates were validated to outperform the best competitive alternatives in the challenging case of crafting multiqubit entangling gates.

We first constructed single-qubit $R_{x}(\pi/2)$ gates, which outperform the IBM default gate in RB with up to a $3\times$ reduction in gate duration and robustness against common system drifts. We then constructed novel implementations of the entangling $ZX(-\pi/2)$ gate which show up to $\approx 2.38\times$ higher fidelity, achieving $\mathcal{F}_{ZX}>99.5\%$.  With these two driven gates, we used randomized benchmarking techniques to validate a complete universal gateset with performance superior to hardware defaults even weeks past last calibration.

From these results We conclude that DRL is an effective tool for achieving error robust gatesets which outperform default, human defined operations by capturing unknown Hamiltonian terms through direct interaction with experimental hardware, and without the need for onerous Hamiltonian tomography methods. Moreover, we have validated that even in the face of restricted access to measurement data, DRL can effectively design useful novel controls.  We expect that in circumstances allowing better access to measurement data, the richness of DRL may allow the construction of gate implementations which are out of reach for simpler cost function minimization methods.

Looking forward, we believe these results validate DRL's utility for directly improving the performance of small-to-medium-scale algorithms, beyond individual gate operations. For instance, it may be beneficial to directly optimize frequently employed circuit elements outside of the underlying universal gateset~\cite{Shi_2019, Gokhale_2019}. Our early experimental exploration of the SWAP gate suggests that additional optimization benefits may be achieved through autonomous gate optimization \textit{in situ}, in order to effectively capture additional transients and context dependent error sources that arise at the circuit level. We look forward to future work extending the applicability of DRL to deliver further algorithmic advantages across a variety of quantum computing applications.  

\begin{acknowledgments}
We acknowledge the IBM Quantum Startup Network for provision of hardware access supporting this work. The views expressed are those of the authors, and do not reflect the official policy or position of IBM or the IBM Quantum team. The authors also acknowledge N. Earnest-Noble for technical discussions and his support enabling our experiments. The authors are grateful to all other colleagues at Q-CTRL whose technical, product engineering, and design work has supported the results presented in this paper.
\end{acknowledgments}

\appendix
\section{Methods}
In this section we briefly summarize the parameters and procedures we used to produce the results in the main text.

\textbf{Reward (cost) Functions for Single Qubit Gate} -- 
In order to evaluate the complete gate performance we repeat the candidate implementation of $R_x(\pi/2)$ a variable number of times $r_{i}$. First, we perform a full state tomography in the computational space to find the fidelity with respect to the ideal target state $\mathcal{F}_{r_{i}}^\text{(qubit)}$. We then calculate the population of the second level, $\ell = P(|2\rangle)$, and re-scale the fidelity
$\mathcal{F}_{r_{i}}^{\text{(qutrit)}} = \mathcal{F}_{r_{i}}^{\text{(qubit)}}\Big/(1+\ell^2)$. 
The reward function is then calculated as a weighed mean of the different repetitions (Fig.~\ref{fig:rl_scheme}d-e).

\textbf{Single Qubit RB} -- 
For the single qubit case we used a customized RB module which generates the $24$ single-qubit Clifford gates using only virtual Z-rotations together with a given $R_x(\pi/2)$ (optimized or default) and construct arbitrary RB sequences out of these gates. The data in the main text consist $18$ sequence lengths up to a maximal sequence length of $2280$ Clifford gates. For each sequence length we generated $20$ random sequences and repeated each $1024$ times in order to estimate the survival probability. The mean (over random sequences) of the survival probability $F$ was then fitted against the sequence length $m$ to the following functional form: $F = A\alpha^m + B$. Since the error per gate is related to the $\alpha$ parameter and since the $A,B$ parameters capture device effects such as SPAM, we first fit all three parameters, both for the default and for the optimized pulse. Then, we re-fit for $\alpha$ with fixed values of $A$ and $B$ which we set to the mean of the unconstrained values. The error per Clifford is then given by $r=(1-\alpha)/2$ and the error per gate is $6r/7$ as the chosen set of the $24$ single-qubit Clifford gates contains $28$ appearances of $R_x(\pi/2)$, meaning $7/6$ $R_x(\pi/2)$ per Clifford. 

\textbf{Two Qubit RB/IRB} -- 
For evaluating two qubit gates we employ the IBM module both for RB and IRB. 
The IRB procedure involves comparing two survival-probability decay curves (each averaged over randomizations) in order to extract an effective EPG for the target CNOT in isolation ~\cite{Magesan2012}. The RB protocol only uses default gates while IRB interleaves a target gate under test with default-implemented Clifford gates. 
The data in the main text consists of $10$ sequence lengths up to a maximal sequence length of $90$ Clifford gates for the CNOT testing and up to $65$ Clifford gates for the SWAP testing. Similar to the single qubit case, for each sequence length we generated $20$ random sequences and repeated each $1024$ times in order to estimate the survival probability. Similar fitting technique was used for the IRB data in order to estimate the relevant $\alpha$ parameter. 

\textbf{Repetition Based Experiments} -- In these experiments we fit the mean infidelity vs. the number of gate repetitions $N$, separate from the repetitions employed in evaluating the reward function. For each value of $N$ we average over $5$ experiments applying the gate under test $N$ times on one initial state, and repeat for a different initial state. For the $ZX$ gate testing the initial states were $|00\rangle$ and $|10\rangle$ and for the SWAP gate testing $|+0\rangle$ and $|+1\rangle$. After each run, a full state tomography was performed and the infidelity with respect to the ideal target state was calculated. An effective measure of error-per-gate is extracted by applying a linear fit to the average infidelity as a function of $N$, which provides a measure to gate error in the low error limit (as cross-validated using IRB). 

\section{Reinforcement Learning Algorithm}
The DRL algorithm used for the gate optimizations in this paper is an on-policy algorithm from the policy gradient family with a stochastic policy and a discrete action space. 
It is a variant of the well known Monte-Carlo policy gradient algorithm, REINFORCE~\cite{williams1992simple}. 
A parameterized policy $\pi_\theta$ is iteratively updated to maximize the discounted episodic return, $J(\theta) = \mathbb{E}_{\tau \sim \pi_\theta(\tau)}[R(\tau)]$. 
It does so by directly estimating the objective's gradient with respect to the policy parameters $\theta$, and then performs a policy update using the Adam~\cite{kingma2015} optimization algorithm, which was chosen due to its overall effectiveness in dealing with non-convex and slowly changing objective landscapes.

It can be shown that 
\begin{align} \label{eqn:J}
\nabla_\theta J(\theta) 
    &= \mathbb{E}_{\tau \sim \pi_\theta(\tau)}[\nabla_\theta \log \pi_\theta(\tau) R(\tau)]\\
    &= \mathbb{E}_{\tau \sim \pi_\theta(\tau)} \bigg[ \sum_{k=1}^{N_{seg}} \nabla_\theta \log \pi_\theta(a_k | s_k) R(\tau)\bigg],
\end{align}
where $R(\tau)$ holds the total discounted return for an episode under trajectory $\tau$. 
This expectation can be efficiently estimated by averaging over a batch of concurrent episodes. 
This overall learning process has the advantage of being straightforward, not requiring nor forming a model of the learning environment, and having sufficient computational efficiency to be effective for gate optimization.

The agent's policy $\pi_\theta$ provides actions which change the agent's state within the environment. 
Each action consists of amplitude and phase values, with a sequence of $N_{seg}$ actions constructing a full PWC control pulse. 
The policy is represented by a feedforward network with one tanh activated hidden layer and a softmax output layer. 
For the single-qubit case, a hidden layer of size 10 was used, and for the two-qubit case, the size was 18.
The softmax output layer provides a probability distribution over the discrete action space, which is then sampled to select the next concrete action to take in the environment.

\begin{figure}[t]
\begin{algorithm}[H] \caption{\hspace{6px} Policy Network Training Step} \label{algo:3}
\begin{algorithmic}[1]
\REQUIRE Batch of $\{\tau_b\}_{b=1}^{B}$ trajectories from latest episode \\
\REQUIRE Policy network parameters: $\theta$ \\
\REQUIRE Discount factor: $\gamma \in [0,1]$ \\
\REQUIRE Current learning rate, and decay rate: $\alpha, \delta$ \\
\FOR {$k = 1,2,\dots,N$} 
\STATE Discount rewards: $R_b[k] = \sum_{i=k}^{N} r_{b,i}\gamma^{N-i}$ \newline for each $b = 1,2,\dots,B$\\
\ENDFOR
\STATE Estimate $\nabla_\theta J \approx \frac{1}{B} \sum_{b=1}^{B} \nabla \log(\pi_\theta(\tau_b)) \cdot R_b$ as Eqn.~\ref{eqn:J}
\STATE Perform an Adam update step on $\theta$ using $\nabla_\theta J$
\IF{$\alpha > \alpha_{\text{min}}$}
\STATE Perform learning rate decay: $\alpha \leftarrow \delta\alpha$
\ENDIF
\end{algorithmic} 
\end{algorithm}
\end{figure}

The agent's policy is updated at the end of each episode, using a batch of trajectories collected throughout the episode. 
The training step is summarized in Alg.~\ref{algo:3}, and is considered \textit{on-policy}, since the actions used for the update were generated by the agent using its current policy, as opposed to a previous policy or an $\epsilon$-greedy version of $\pi_\theta$. 
In our use case, a trajectory $\tau$ consists of a sequence of pulse segments applied, the tomographic state measurements, and the fidelity based reward received after completion of the pulse construction at the conclusion of an episode.

The policy updates work to maximize Eqn.~\ref{eqn:J}. 
Practically, these updates are determined by computing the policy network loss, which is the negative cross-entropy between the predicted probabilities for each possible action and the chosen actions throughout the episode, weighted by the episode's discounted rewards. 
This is then minimized using the Adam optimizer using default parameters. 
By minimizing the loss, or equivalently maximizing the log-likelihood, the network is encouraged to assign higher probabilities to actions which previously led to larger episodic returns.

\section{Fast Simulated Annealing}
\begin{figure}[t!] 
\centering
\includegraphics[width=\linewidth]{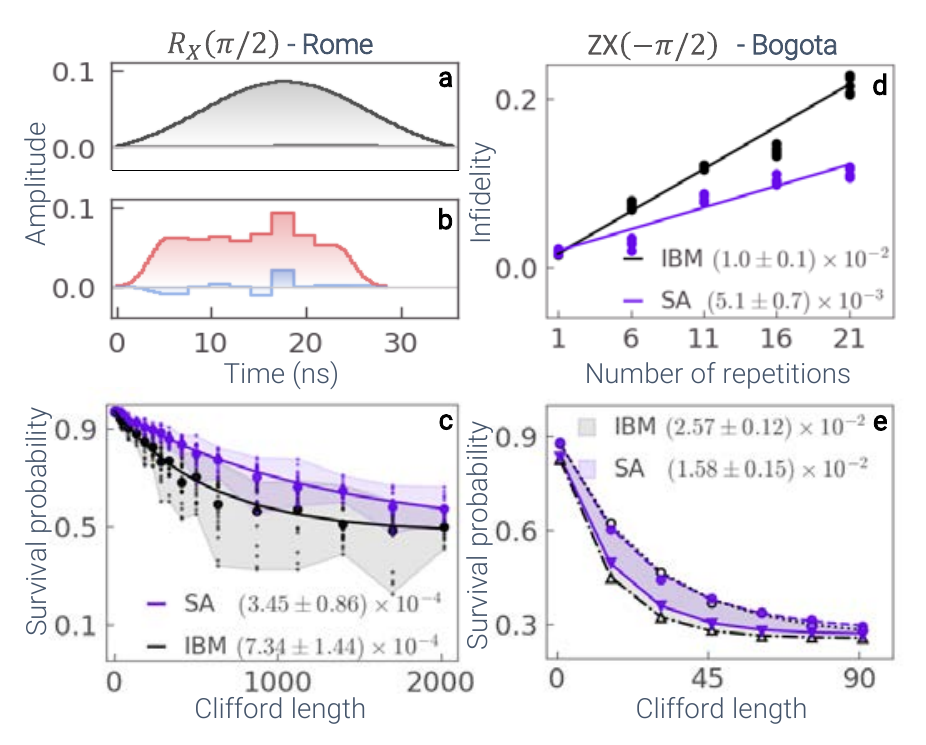}
\caption{Additional data on Simulated Annealing closed-loop optimization. (a-c) Optimization of $R_x(\pi/2)$ gate on \textit{ibmq\_rome}. The optimized pulse (b) is $20\%$ shorter than the IBM default (a) and has $2.13\times$ lower gate error as measured using RB (c).  (d-e) Optimization of $ZX(-\pi/2)$ and composite CNOT on \textit{ibmq\_bogota}. Both repetitions (d) and IRB (e) show $\sim2\times$ improvement in the gate error compared to the default gate, consistent with previous data sets appearing in the main text.  Absolute error rates for the \textit{ibmq\_bogota} device appeared consistently higher than other machines tested.}
\label{fig:SA}
\end{figure}

In the main text we explored the performance of an automated Fast Simulated Annealing algorithm, Cauchy machine, in optimizing a two-qubit gate on the IBM machine. Here we provide details about the SA optimization process and present additional results of an $R_x(\pi/2)$ gate optimization on \textit{ibmq\_rome} and an optimization of $ZX(-\pi/2)$ on \textit{ibmq\_bogota}. The results of the optimization processes appear in Fig.~\ref{fig:SA}.

For the SA optimization process, no intermediate information is collected and the evaluation of the full gate performance is performed with the same reward function we used in the RL optimization to estimate the full gate implementations, \textit{i.e.}, a weighed mean of the state fidelities. 
The starting point of the SA algorithm is the device default for the gate we wish to optimize. The general SA optimization process is summarized in Alg.~\ref{algo:2}.

\begin{figure}[H]
\begin{algorithm}[H] \caption{\hspace{6px} SA Training Loop} \label{algo:2}
 \begin{algorithmic}[1] 
 \REQUIRE {Initialize temperatures $T_0^\text{cost}$, $T_0^\text{amp}$, $T_0^\text{phase}$}
 \REQUIRE {Initialize amplitudes ($A_i$) and phases ($\phi_i$) to the default values }
  \FOR {step $= 1,2,\ldots$}
   \STATE Draw an amplitude scale: $\delta A=Ch(0, T^\text{amp})$ \COMMENT{$Ch$ is a Cauchy distribution}
   \STATE Draw a phase scale: $\delta\phi=Ch(0, T^\text{phase})$ 
   \STATE Draw two unit vectors $u_i$, $v_i$
   \STATE Shift the amplitudes $A^\text{temp}_i = A_i+ u_i \delta A$
   \STATE Shift the phases $\phi^\text{temp}_i = \phi_i+ v_i \delta \phi$
   \STATE Calculate candidate cost $C=\text{cost}(A^\text{temp},\phi^\text{temp})$
   \IF {$C < C_\text{best}$}
   \STATE Accept candidate
    \ELSE{} 
    \STATE Accept candidate with probability $F\big(\frac{C_\text{best} - C}{T^\text{cost}}\big)$ \COMMENT{$F$ is the acceptance distribution}
    \ENDIF
    \STATE Update the three temperatures: $T = \frac{T_0}{1+\text{step}}$
   \ENDFOR
 \end{algorithmic}
\end{algorithm}
\end{figure}
\vspace{-24px}


%
\end{document}